\documentclass[prl,superscriptaddress,twocolumn,preprintnumbers,showkeys,showpacs,floatfix]{revtex4}
\usepackage{url}
\usepackage{epsf}
\usepackage{amsfonts,amsmath,amsbsy,latexsym}

\newcommand{\eM}     {$\epsilon$-machine}
\newcommand{\eMs}    {$\epsilon$-machines}

\newcommand{\EMs}    {$\epsilon$-Machines}

\newcommand{\Abet}              { {\cal A} }

\newcommand{\CausalState}       { {\cal S} }

\newcommand{\CausalStateSet}    { \boldsymbol{\CausalState} }
\newcommand{\TransProbSet}      { \boldsymbol{\cal T} }

\newcommand{\Prob}              { {\rm P} }

\newcommand{\Cmu}               { C_\mu }

\newcommand{\Alphabet}           { \mathcal{A} }

\begin{document}

\markboth{Objects that make Objects}{J. P. Crutchfield and O. G\"{o}rnerup}

\title{Objects That Make Objects:\\
The Population Dynamics of Structural Complexity}

\author{James P. Crutchfield}
\email[Electronic address: ]{chaos@santafe.edu}
\affiliation{Santa Fe Institute, 1399 Hyde Park Road, Santa Fe, NM 87501}
\author{Olof G\"{o}rnerup}
\email[Electronic address: ]{olof@santafe.edu}
\affiliation{Santa Fe Institute, 1399 Hyde Park Road, Santa Fe, NM 87501}
\affiliation{Department of Physical Resource Theory, Chalmers
University of Technology \& G\"{o}teborg University, 412 96 G\"{o}teborg, Sweden}

\date{\today}

\begin{abstract}
To analyze the evolutionary emergence of structural complexity in physical
processes we introduce a general, but tractable, model of objects that
interact to produce new objects. Since the
objects---\emph{$\epsilon$-machines}---have well defined structural
properties, we demonstrate that complexity in the resulting population
dynamical system emerges on several distinct organizational scales during
evolution---from individuals to nested levels of mutually self-sustaining
interaction. The evolution to increased organization is dominated
by the spontaneous creation of structural hierarchies and this, in turn, is
facilitated by the innovation and maintenance of relatively low-complexity,
but general individuals.
\end{abstract}


\preprint{Santa Fe Institute Working Paper 04-06-020}
\preprint{arxiv.org e-print adap-org/0406XXX}

\maketitle



Long before the distinction between genetic information and functional
molecules---molecules encoded by, but manipulating that stored
information---there were objects that simply interacted and mutually
transformed each other. How did structured objects with the dual roles
of information storage and transformation emerge in such initially disorganized
environments? Here we introduce a class of models that allows us to explore
this question, in a setting often referred to as \emph{pre-biotic} evolution
\cite{Rasm04a}.
In contrast with prior work, we focus on the questions of what \emph{levels}
of structure and information processing can emerge and, specifically, what
population-dynamical mechanisms drive the transition from pre-biotic
to biotic organization.

One of the key puzzles in this is to understand how systems, on the one hand,
produce structure that, on the other, becomes substrate for future functioning
and innovation. The spirit of our approach to this puzzle follows that
suggested by Schr\"odinger \cite{Schr67a} and found in von Neumann's random
self-assembly model \cite{Neum66a}. However, our model is more physical than
chemical in the sense that we do not assume the existence of sophisticated
chemical entities, such as macromolecules, nor do we even use chemical
metaphors, such as information being stored in one-dimensional arrays---in
\emph{aperiodic crystals}, as Schr\"odinger presciently proposed. While
ultimately interested in pre-biotic organization and its emergence, our focus
is on what one might call \emph{pre-chemical} evolution. As such, the
following provides a first step to directly address how structural complexity
and evolutionary population dynamics interact \cite{Crut01d}.

Here we introduce a model of the emergence of organization and
investigate it in a setting that, at one and the same time, provides a
well defined and quantitative notion of structure and is mathematically
tractable. The goal is to develop predictive theories of the population
dynamics of interacting, structured individuals and their collective
organization. With a well defined measure of structural complexity one
can precisely state the question of whether or not complexity has genuinely
emerged over time in pre-biotic and pre-chemical processes. Additionally,
with a predictive theory of the population dynamics one can identify and
analyze the (at some point evolutionary) mechanisms that lead to such
increases (and to decreases) in structural complexity.

We create a well stirred population---the \emph{finitary process soup}---of
initially random, finite \eMs\ \cite{Crut88a} that interact and transform
each other, making new \eMs\ of differing structure and so of differing
transformational properties. The initial random soup serves as a reference
that, for reasons to become apparent below, has ``null'' structural complexity,
both in the individuals (on average) and across the population. Here we
consider the case of a \emph{process gas}: objects, \eMs\ $T_A$ and $T_B$,
are successively randomly paired (\emph{pan-mixia}) and act on each other
to create progeny: $T_B + T_A \rightarrow T_C$. No externally applied
selection or variation is imposed.

An \emph{\eM}\ $T = \{\CausalStateSet, \TransProbSet\}$ consists of a set of
\emph{causal states} $\CausalStateSet$ and \emph{transitions} $\TransProbSet$
between them: $T_{ij}^{(s)}, ~s \in \Abet$. We interpret the symbols labeling
the transitions in the alphabet $\Abet$ as consisting of two parts: an
\emph{input symbol} that determines which transition to take from a state and
an \emph{output symbol} which is emitted on taking that transition. \eMs\
have several key properties \cite{Crut88a}: All of their recurrent states form
a single \emph{strongly connected} component. Their transitions are
\emph{deterministic} in the specific sense that a causal state together with
the next input symbol determines the successor state. And, $\CausalStateSet$
is \emph{minimal}: an \eM\ is the smallest causal representation of the
transformation it implements.

Due to these properties, one can quantify an \eM's structural complexity. To
do this we need the probability distribution over the states in
$\CausalStateSet$, how often they are visited, and this is given by the
normalized left eigenvector associated with eigenvalue $1$ of the stochastic
matrix $\mathbf{T}\equiv \sum_{s\in \Alphabet} T^{(s)}$. Denote this
eigenvector, normalized in probability, by $p_{\CausalStateSet}$. An \eM's
\emph{structural complexity} is the amount of stored information:
$\Cmu (T) \equiv -\sum_{v\in \CausalStateSet} p_{\CausalStateSet}^{(v)} \log_2 p_{\CausalStateSet}^{(v)}$.
When $\Cmu$ is finite, we say the \eM\ (or, more properly, the transformation
it describes) is \emph{finitary}.

Thus, unlike previous models---such as $\lambda$-expressions \cite{Font91a},
machine instruction codes \cite{Ray91a}, tags \cite{Bagl89a}, and cellular
automata \cite{Crutchfield&Mitchell94a}---\eMs\ allow one to readily
measure the structural complexity and disorder of the transformations they
specify. It is well known that algorithms do not even exist to measure these
quantities for machine-language programs and $\lambda$-expressions, for example,
since these are computation-universal models \cite{Broo89a}. As our results
demonstrate, these tractable aspects of \eMs\ give important quantitative,
interpretive, and theoretical advantages over prior work on the pre-biotic
evolutionary emergence of structural complexity. The finitary assumption is
also consistent with the recent proposal that gene expression is implemented
with finite-memory computational steps \cite{Bene04a}.

We should emphasize that, in these finitary process soups and in
contrast with
prior work, \eMs\ do not have two distinct modes of representation or
functioning. The objects are only functions, in the prosaic mathematical sense.
Thus, one benefit of this model of pre-biotic evolution is that there is no
assumed distinction between gene and protein \cite{Schr67a,Neum66a}, between
data and program \footnote{One recovers the dichotomy by projecting onto (i)
the sets that an \eM\ recognizes and generates and (ii) the mapping between
these sets \cite{Broo89a}.}.

Finitary process soups allow one to quantitatively analyze not only the
structural complexity of individuals, but also the interaction between
individual structure and population organization and dynamics in terms of
how they store and process information and the \emph{causal architecture}
that supports these. Since this view of a system {\em intrinsically computing}
applies both to individuals and to the population as a whole, we can
identify the locus of a population's structural complexity. Is it largely
the sum of the individuals' or largely embodied in the transformative
relationships between individuals? Perhaps it derives from some irreducible
interaction between these levels.

The finitary process soup differs from early investigations in which
finite-state machines were evolved using an explicit mutational operator
\cite{Foge66a}. Here, novelty derives directly from how the objects
themselves are structured, since this determines how they transform each other.
Equally important, survivability is determined by an individual's rate of
reproduction---the original biological notion of \emph{fitness}; there is no
externally imposed fitness function. In this, the process soup is similar to
the molecular evolution models of Eigen and Schuster \cite{Schu77a}.

A \emph{population} $P$ is a set of $N$ \emph{individuals}, each of which is an
\eM. More compactly, one can also describe the population as a distribution of
\eM\ types: $\mathbf{f} = (a_1/N, a_2/N, \ldots, a_n/N)$, where $n$ is the
number of possible \eM\ types and $a_i$ is the number of individuals of type
$T_i$. A single replication is determined through compositions and replacements
in a two-step sequence: First, construct \eM\ $T_C$ by:
\begin{enumerate}
\item With probability $1 - \Phi_\mathrm{in}$, forming the composition
	$T_C = T_B \circ T_A$ from $T_A$ and $T_B$ randomly selected from the
	population and minimizing \cite{Broo89a}.
\item With probability $\Phi_\mathrm{in}$, generating a random $T_C$.
\end{enumerate}
Second, replace a randomly selected \eM, $T_D$, with $T_C$. 
$\Phi_\mathrm{in}$ is the rate of influx of new (random) \eMs. When
$\Phi_\mathrm{in} = 0$, the soup is a closed system. When
$\Phi_\mathrm{in} = 1$, the soup is open, but consists of entirely random
\eMs\ and so is unstructured. The initial population $P_0$---with
$\mathbf{f} = (1/n, \ldots, 1/n)$---is similarly unstructured.
 
As a first step to detect population structure we define the
\emph{interaction network} $\mathcal{G}$ as the \eM\ compositions that have
occurred in the population. For a population with $n$ different types,
$\mathcal{G}$ is described by an $n \times n$ matrix the entries of which are
the machine types returned by the compositions $T_i \circ T_j,~ T_i, T_j \in P$.
We represent $\mathcal{G}$ as a graph whose nodes are the machine types in the
population and whose directed edges connect one node, say $T_A$, to another,
$T_C$, when $T_C = T_B \circ T_A$. The edges are labeled with the transforming
machine $T_B$. We also represent $\mathcal{G}$ as a transition matrix
$\mathcal{G}_{ij}^{(k)} = \Prob (T_k | T_i , T_j)$, when $T_k = T_j \circ T_j$.

The second step is to introduce a natural notion of organization that
encompasses interaction and dynamic stability, we define a \emph{meta-machine}
as a set of \eMs\ that is both closed and self-maintained under composition.
That is, $\Omega \subset P$ is a meta-machine if and only if (i)
$T_i \circ T_j \in \Omega$, for all $T_i, T_j \in \Omega$ and (ii) for all
$T_k \in \Omega$, there exists $T_i, T_j \in \Omega$, such that
$T_k = T_i \circ T_j$. This definition of self-maintenance captures
Maturana et al's \emph{autopoiesis} \cite{Vare74a}, Eigen and
Schuster's \emph{hypercycles} \cite{Schu77a}, and \emph{autocatalytic sets}
\cite{Kauf86a,Font91a}. In a process soup, awash in fluctuations and change,
a meta-machine is a type of organization that can be regarded as an autonomous
and self-replicating entity. Note that, in this sense, the initial random soup
$P_0$ is not organized. To the extent that interaction networks persist, they
are meta-machines.

To measure the diversity of interactions in a population we define the
\emph{interaction network complexity}
$\Cmu ( {\mathcal G} ) = - \sum_{p_i, p_j, p_k > 0} v_{ij}^k \log_2 v_{ij}^k$,
where
\begin{equation}
v_{ij}^k = \left\{
  \begin{array}{ll}
  p_i p_j / \sum v_{ij}^k, & T_k = T_i \circ T_j \mathrm{~has~occurred,}\\
  0, & \mathrm{otherwise}.
  \end{array}
  \right.
\end{equation}

Finally, a machine type's frequency changes at each generation according
to its interactions and is given by
\begin{equation}
{\mathbf f}_t^{(k)} =
  \mathbf{f}_{t-1} \cdot \mathcal{G}_{ij}^{(k)} \cdot \mathbf{f}_{t-1}
  / \sum_{k=1}^n
  \mathbf{f}_{t-1} \cdot \mathcal{G}_{ij}^{(k)} \cdot \mathbf{f}_{t-1} ~.
\label{PopulationDynamics}
\end{equation}

Let us now explore a base case: the population dynamics of one simple subset of
\eMs, those consisting of only a single state. This class is especially
instructive since it is closed under composition: the composition of two
single-state machines is itself a single-state machine. There are $15$
single-state \eMs; excluding the null machine. As a consequence, there is a
finite number of possible interactions and this, in turn, greatly facilitates
an initial analysis. Although a seemingly trivial case, a population
of these machines exhibits nontrivial dynamics and leads to several insights
about unrestricted populations.

\begin{figure}[h]
\begin{center}
\epsfxsize=3.4in
\epsffile{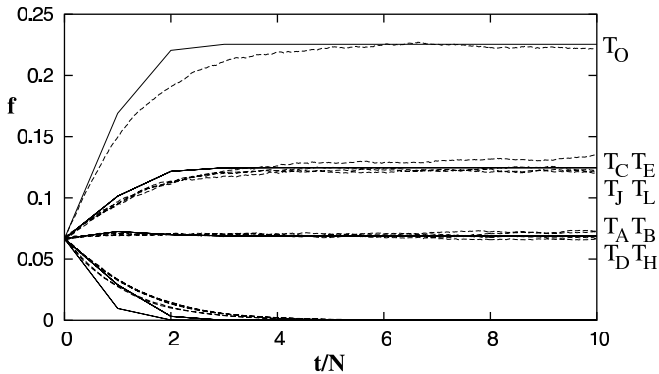}
\caption{Population dynamics starting with $N = 10^5$ randomly sampled
  single-state \eMs: Fraction $\mathbf{f}_t$ of \eM\ types as a function of
  time $t$ (number of replications).
  Simulations (dashed lines) and theory Eq. (\ref{PopulationDynamics})
  (solid lines).
  }
\label{SingleStatePopulationDynamics}
\end{center}
\end{figure}

The first obvious evolutionary pressure driving the system is that governed
by trivial self-reproduction (copying). \EMs\ with the ability to copy
themselves (directly or indirectly) are favored, possibly at other
\eMs' expense or in symbiosis with other \eMs. The number of such
self-reproducing machines grows in relation to the whole population,
further increasing the probability of self-reproduction. Interaction networks
that sustain this will emerge, consisting of cycles of cooperatively
reproducing \eMs. These are chains of composed mappings that form closed loops.

\begin{figure}[h]
\begin{center}
\epsfxsize=3.6in
\epsffile{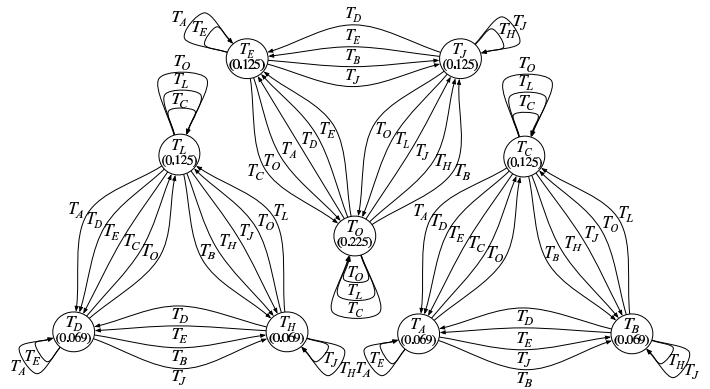}
\caption{Organization of the steady-state population---a meta-machine: The
  interaction graph $\mathcal G$ after $10^6$ replications, at the
  end of Fig. \ref{SingleStatePopulationDynamics}. The graph shows only
  $T_A \stackrel{T_B}{\longrightarrow} T_C$-denoted interactions.
  $\Cmu ( {\mathcal G} ) \approx 5.75$ bits.
  }
\label{SingleStateInteractionNetwork}
\end{center}
\end{figure}

Figure \ref{SingleStatePopulationDynamics} shows the dynamics of a population
sampled from $\mathbf{f}_0 = (1/15, \ldots, 1/15)$ and in a closed system
($\Phi_\mathrm{in} = 0$) of $N = 10^5$ \eMs. The figure shows that Eq.
(\ref{PopulationDynamics}) predicts the simulations quite well. Out of the
interactions between all possible \eMs\ the population settles down to a
steady-state interaction network of nine \eMs. Figure
\ref{SingleStateInteractionNetwork} shows this meta-machine. Note that the
structural complexity of individual \eMs\ is always zero: $\Cmu (T) = 0$
for single-state machines. Thus, the population's structural complexity is
due solely to that coming from the network of interactions.

\begin{figure}[h]
\begin{center}
\epsfxsize=3.5in
\epsffile{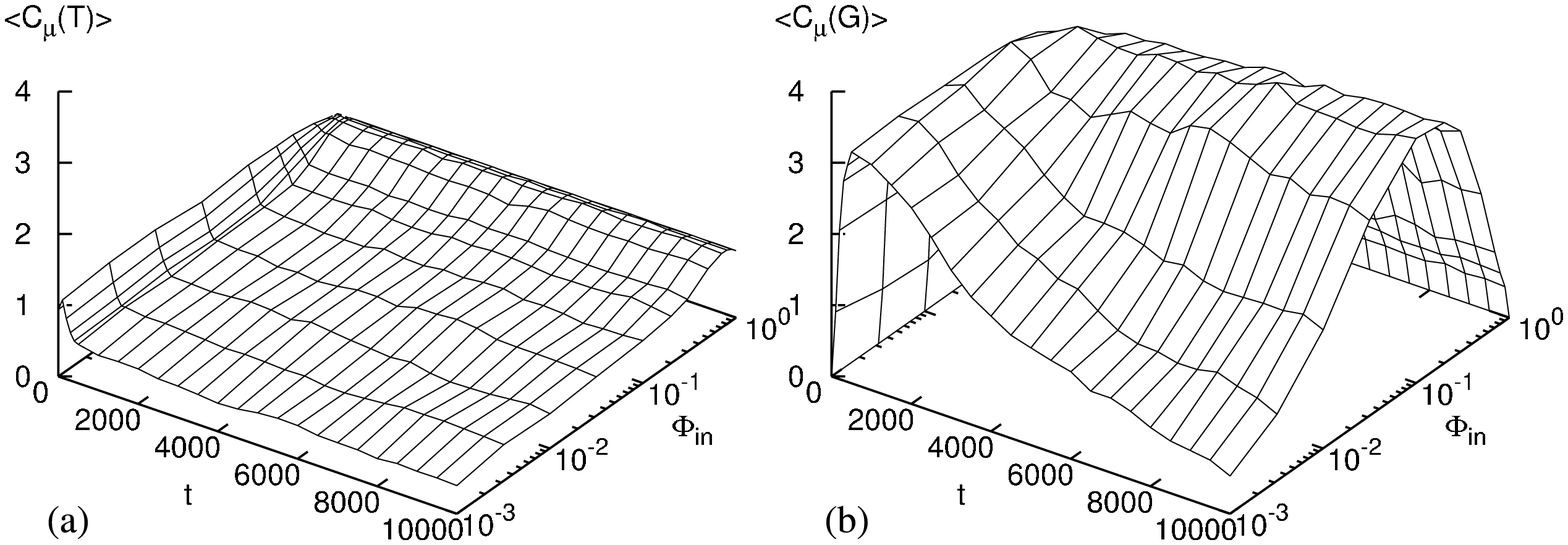}
\caption{(a) Population-averaged \eM\ complexity $\langle \Cmu (T) \rangle$
  [bits] and (b) run-averaged interaction network complexity
  $\langle \Cmu (\mathcal{G}) \rangle$ [bits] versus time $t$ and influx rate
  $\Phi_\mathrm{in}$ for a population of size $N = 100$ averaged over
  $50$ runs at each $\Phi_\mathrm{in}$.
  }
\label{PolyStateInfluxTimeSurface}
\end{center}
\end{figure}

With ways to predict the population dynamics and to detect the emergence of
structural complexity in the soup, we now turn to the evolution of
unrestricted populations. We summarize the results using the
population-averaged \eM\ complexity $\langle \Cmu (T) \rangle$
and the run-averaged interaction network complexity
$\langle \Cmu (\mathcal{G}) \rangle$ as a function of time and influx rate;
see Fig. \ref{PolyStateInfluxTimeSurface}. One observes an initial rapid
construction of increasingly complex individuals and interaction networks.
In the closed system ($\Phi_\mathrm{in} = 0$), both of these reach a maximum
and then decline to less complex steady states within a small subspace of
possible structures. In fact, both structural complexities effectively vanish
at this extreme. The closed system specializes, ages, and eventually dies away.
At the extreme of high influx ($\Phi_\mathrm{in} = 1$), when the population
looses the ability to store information, the network complexity vanishes and
the individual complexity becomes that of a purely random sample of \eMs.

Away from these extremes, the evolution of the open systems' network complexity
is maximized at an intermediate influx rate $\Phi_\mathrm{in} \approx 0.10$.
Notably, the emergence of complex organizations occurs where individual
\eM\ complexity is small. Survival, however, requires these individuals to
participate in interaction networks and so to interact with a variety of other
machines; they are generalists in this sense. At higher influx
($\Phi_\mathrm{in} \approx 0.75$) large $\langle \Cmu (T) \rangle$ is
correlated with markedly less complex networks. These more complex machines
are specialized and do not support robust complex interaction networks.

It turns out that the maximum network complexity $\widehat{\Cmu} (\mathcal{G})$
grows slowly (linearly) with time. It is ultimately capped by the population
size since there is only so much structure that can be built with a finite
number of components. More extensive investigations show that it grows in an
unbounded way---$\widehat{\Cmu} (\mathcal{G}) \propto \log N$---indicating the
possibility of reaching highly structured populations at large sizes.

The finitary process soup demonstrates (i) that complexity of the entire
system arises mainly from the transformative relationships between individuals
and (ii) that those individuals tend to be noncomplex and to implement
general rather than specialized local functions. Thus, the population dynamics
makes a trade-off: simpler individuals facilitate the emergence of global
structure. Conversely, for a system to become complex, it is not necessary
to evolve complex individuals. The results strongly suggest that replicative
processes will use this particular strategy to build successively higher
levels of structural complexity from the compositional (``metabolic'') network
of interacting finitary components. In this way, the finitary process soup
evolves higher computational representations, they are not built-in or
accessible at the outset.

The finitary process soup is a model of endogenous evolution. In particular,
fitness is defined and measured in the biologically plausible way, as the
rate of individual reproduction, and there is no externally imposed mutational
operator. It is also a flexible model; we showed pan-mixia replication and
will report on populations with spatial structure elsewhere. Moreover, it is
extensible in a number of ways. Specifically, it is straightforward to couple
in energetic and material costs of composition. This will allow one to analyze
trade-offs between energetics, dynamics, and organization. Finally, one of the
key determinants of evolutionary dynamics is the structure of selection-neutral
genotype networks; see Ref. \cite{Crut01d} and references therein. The neutral
networks of \eMs\ and of \eM\ networks can be directly probed.

We close with a few general comments. Given that organization in a population
becomes hierarchical, we believe that powerful computational representations,
when employed as the \emph{basic} objects, are neither effectively used by
nor necessary for natural evolutionary processes to produce complex organisms.
We hypothesize that individuals with finitary computational capacity are
appropriate models of molecular entities and transformations. From these, more
sophisticated organizations and functions can be hierarchically assembled. This
can only be tested by experiment, of course, but this will soon be possible.

It has been recently estimated that the genomes of higher species consist
of a surprisingly small number of genes compared to the number found in lower
species \cite{Lync03a,RGSPC04a}, despite the higher species being markedly
more complex and diverse in their behaviors. Moreover, many of those genes
serve to maintain elementary functions and are shared across species. These
observations accord with the evolutionary dynamics of the finitary process
soup: global complexity is due to the emergence of higher level structures
and this in turn is facilitated by the discovery and maintenance of
relatively noncomplex, but general objects. In both the genomic and
finitary soup cases, one concludes that an evolving system's sophistication,
complexity, and functional diversity derive from its hierarchical
organization.

This work is supported by Intel Corporation, core grants from the National
Science and MacArthur Foundations, and DARPA Agreement F30602-00-2-0583.
OG was partially supported by the International Masters Programme in Complex
Adaptive Systems.

\vspace{-0.25in}
\bibliography{chaos}

\end{document}